Magnetic Fields Threading Black Holes:
restrictions from general relativity and implications for astrophysical black holes

David Garofalo[1]
1. Department of Physics, Kennesaw State University, Marietta GA 30060

Abstract

The idea that black hole spin is instrumental in the generation of powerful jets in active galactic nuclei and X-ray binaries is arguably the most contentious claim in black hole astrophysics. Because jets are thought to originate in the context of electromagnetism, and the modeling of Maxwell fields in curved spacetime around black holes is challenging, various approximations are made in numerical simulations that fall under the guise of 'ideal magnetohydrodynamics'. But the simplifications of this framework may struggle to capture relevant details of real astrophysical environments near black holes. In this work, we highlight tension between analytic and numerical results, specifically between the analytically derived conserved Noether currents for rotating black hole spacetimes and the results of general relativistic numerical simulations (GRMHD). While we cannot definitively attribute the issue to any specific approximation used in the numerical schemes, there seem to be natural candidates, which we explore. GRMHD notwithstanding, if electromagnetic fields around rotating black holes are brought to the hole by accretion, we show from first principles that prograde accreting disks likely experience weaker large-scale black hole-threading fields, implying weaker jets than in retrograde configurations.

1. Introduction

The advent of general relativistic simulations (Koide et al 1998; Koide et al 2000; De Villiers & Hawley 2003; De Villiers, Hawley & Krolik 2003; Gammie et al 2003; Komissarov 2004; De Villiers et al 2005; McKinney & Gammie 2004; McKinney 2006) of electromagnetic fields threading black holes required about two decades from the seminal analytic exploration of Blandford & Znajek (1977; henceforth BZ). But the BZ solution and numerical simulations have for the most part parted ways. For example, the BZ notion that the entire region outside the black hole and accretion disk behaves as though gravitational forces are negligible (the so-called force-free assumption), is not generally accepted. While simulations find that force-freeness in the polar region near black holes seems reasonable, plasma inertia becomes important elsewhere (Hirose et al 2004; however see Komissarov 2004). Despite the fact that numerical simulations have made our exploration of black hole magnetospheres more realistic, there are crucial assumptions that may be problematic. Among these is the idea that the initial electromagnetic field within the accretion disk is independent of any detailed properties of the disk. In other words, in the initial setup for the simulations, conditions must be imposed on the electromagnetic field, but the nature of this setup is difficult to determine because we do not understand how these fields make their way into the inner disk region in the first place. Therefore, initial conditions are employed that may not be compatible with the physical principles that are operative in such environments. The assumed initial conditions may be in conflict with the way that the fields originate near the hole either by dynamo enhancement (Livio 2001; Beckwith 2011) or by inwards advection from elsewhere (Lubow et al 1994; Heyvaerts et al 1996). This question of the advection versus the dynamo origin of the large-scale electromagnetic field threading the black hole remains unresolved. In addition to the issue of the initial conditions on the field, GRMHD simulations solve what are known as the 'ideal MHD' equations in curved spacetime, representing approximations that likely break down in the violent environment near black holes. Perhaps most importantly, however, GRMHD simulations are constrained in their treatment of the microscopic physics due to limitations in resolution. It is to these issues that we will appeal to in trying to understand the tension that emerges between basic principles and GRMHD results.

Our primary goal in this paper, however, is to derive constraints on the strength of the black hole-threading electromagnetic field from first principles. Because the energy allotted to the electromagnetic field comes from the total reservoir of energy available to the accreting black hole, the ways that accreting black holes can transfer energy is not arbitrary, but constrained by the specifics of the accretion process. As mentioned, the large-scale field is most likely either created in-situ by a disk dynamo, advected inwards from elsewhere, or some combination of the two processes. Under the assumption that these are the processes that determine the large-scale field, we identify constraints within general relativity on the energy transferrable to the large-scale field, and that such constraints fit uncomfortably within the picture emerging from GRMHD simulations.

In section 2 we appeal to Noether's theorem to obtain the conserved energy at infinity for an accretion disk around a rotating black hole in Kerr spacetime and derive the black hole spin constraints. These constitute well known results that when coupled to our emerging understanding of how accretion operates, produce powerful constraints. In section 3 we discuss the implications for numerical simulations. In section 4 we summarize and conclude.

2. A Kerr black hole: Noether current and energy conservation

From the recognition that

$$\mathcal{L}_K g_{ab} = 0 \qquad (1)$$

Where $\mathcal{L}$ is the Lie derivative, $g_{ab}$ is the metric, and $K^a$ is a vector field,

$$K_{b;a} + K_{a;b} = 0, \qquad (2)$$

which, via Noether's theorem, implies the following conservation law

$$(K_b T^{ab})_{;a} = 0. \qquad (3)$$

If $K^a$ is the Killing vector associated with time translation,

$$C^a = K_b T^{ab} \tag{4}$$

is a conserved 4-vector corresponding to the flux of energy measured at radial infinity. Since

$$T^{ab} = T^{ab}_{matter} + T^{ab}_{E\&M} \tag{5}$$

where $T^{ab}_{matter}$ is the matter part of the stress-energy tensor and $T^{ab}_{E\&M}$ is the electromagnetic part of the stress-energy tensor, and the fact that

$$C^a = K_b(T^{ab}_{matter} + T^{ab}_{E\&M}), \tag{6}$$

the time-independence of the Kerr metric gives us a conserved energy with an exchange of that conserved energy between $T_{matter}$ and $T_{E\&M}$ but with the total remaining constant. This is all well understood. However, we now show that the mechanism providing the black hole with its large-scale electromagnetic field requires energy to be transferred from matter to electromagnetic fields in a way that depends on the value of the spin of the black hole, thus providing us with constraints on the strength of large-scale electromagnetic fields around rotating black holes. If the physical system under consideration involves the accretion of matter about a rotating black hole operating via the magnetorotational instability (MRI) for the transfer of energy and angular momentum, then the Killing vector $K^a$ can be used to construct a conservation law for the circular geodesic orbits of the accretion disk. More specifically, we build on the analysis of Garofalo (2009) who evaluates

$$K_b p^b \tag{7}$$

and

$$L_b p^b \tag{8}$$

in Boyer-Lindquist coordinates, with $p^a$ and $L^a$ the 4-momentum vector for circular geodesics in the accretion disk and the Killing vector associated with azimuthal invariance of the Kerr metric, respectively, showing that a factor of 60 greater work (in the sense of a work-kinetic energy theorem) is needed in the high prograde regime compared to the high retrograde accretion regime. In other words, in order to produce a fixed accretion rate, the high prograde accreting disk will have to place 60 times as much of the conserved energy in electromagnetic form in the accretion disk compared to the high retrograde case. That analysis is done for the inner disk and therefore ignores the fact that retrograde and prograde disks live in different regions of the gravitational potential of the black hole, a crucial recognition. We generalize that result and show that it is true regardless of the region of the disk that is taken into analysis in the sense that it is true for the entire disk. The focal point in the current analysis, however, will be that while a greater total energy is available in the prograde regime, that additional energy is unlikely to end up in the large scale electromagnetic field in the proportions needed to thread prograde accreting black holes with stronger electromagnetic fields. This analysis occurs in Section 3.

Working with equations (7) and (8) which give us the conserved energy and angular momentum for circular geodesic orbits in the accretion disk, in units where c=G=$M$=1 ($M$ the black hole mass, $a$ the black hole spin, and $r$ the radial coordinate) we obtain the following equations for prograde and retrograde disks (Bardeen et al 1972).

$E_{pro} = (r^{3/2} - 2r^{1/2} + a)/(r^{3/4}(r^{3/2} - 3r^{1/2} + 2a)^{1/2}$

$E_{retro} = (r^{3/2} - 2r^{1/2} - a)/(r^{3/4}(r^{3/2} - 3r^{1/2} - 2a)^{1/2}$

$L_{pro} = (r^2 - 2ar^{1/2} + a^2)/(r^{3/4}(r^{3/2} - 3r^{1/2} + 2a)^{1/2})$

$L_{retro} = (r^2 + 2ar^{1/2} + a^2)/(r^{3/4}(r^{3/2} - 3r^{1/2} - 2a)^{1/2})$

From these we can determine the energy in electromagnetic form that is required in the accretion disk to produce the extraction of a fixed amount of angular momentum by evaluating differences in energy and angular momentum for specified regions of the disk. Our analysis does not require any complex computational strategies and can easily be reproduced. Equations (4) and (5) tell us that the conserved energy in the accretion disk amounts to a combination of matter and electromagnetic contributions. Therefore, as a greater amount of the total energy goes into electromagnetic form in order to ensure that the necessary angular momentum is extracted as determined by equation (8), less of that total is available for largescale electromagnetic fields threading the accretion disk. This is shown in Figure 1, which tells us how this process depends on black hole spin and the orientation of the accretion disk. In particular, note the difference at the extremes of disk orientation, near the high spin regime. There is about a factor of 2.2 difference in energy. Because electromagnetic field energy goes like fields squared, largescale electromagnetic fields threading high spinning prograde accreting black holes should suffer a drop of about a factor of 1.48.

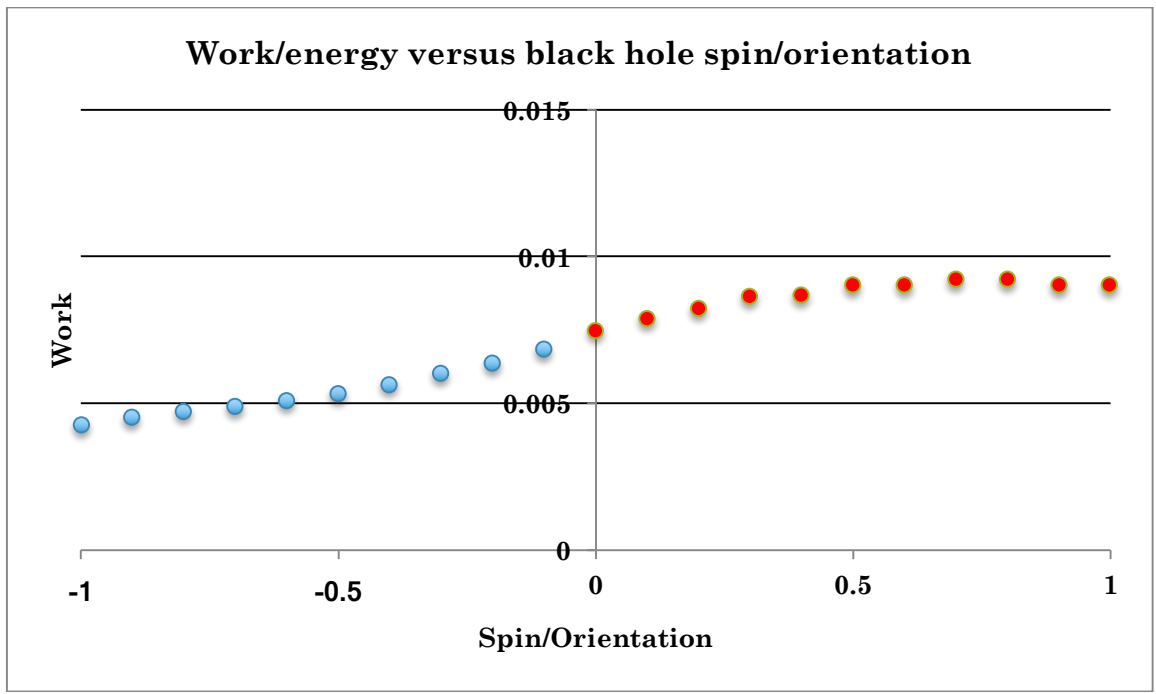

Figure 1: Work versus spin required to produce a fixed accretion rate needed to extract a dimensionless angular momentum of 0.15 from the disk. The amount of electromagnetic energy required as input into the accretion disk at high prograde spin (far right) is about 2.2 times larger than at high retrograde spin (far left).

It is important to emphasize that such a result comes from first principles and is therefore independent of the precise character of spatially connected plasma. Therefore, whether one envisions a linear MRI-generated accretion or a turbulent MRI-based one, the basic fact that spatially separated regions are electromagnetically coupled is subject to our conservation law constraint. In other words, the constraint is independent of the turbulent degree of the accretion disk.

In Figure 2, we show the results obtained from relaxing the fixed angular momentum constraint and evaluating the work required to accrete matter in a thin disk with fixed outer boundary at 30 gravitational radii. While the total amount of angular momentum extraction is different for different black hole spins and orientations of the disk, the trend remains unchanged, namely that a greater work is required to produce an accretion event as the spin increases in the prograde direction.

In the next section we explore this constraint in the context of GRMHD simulations.

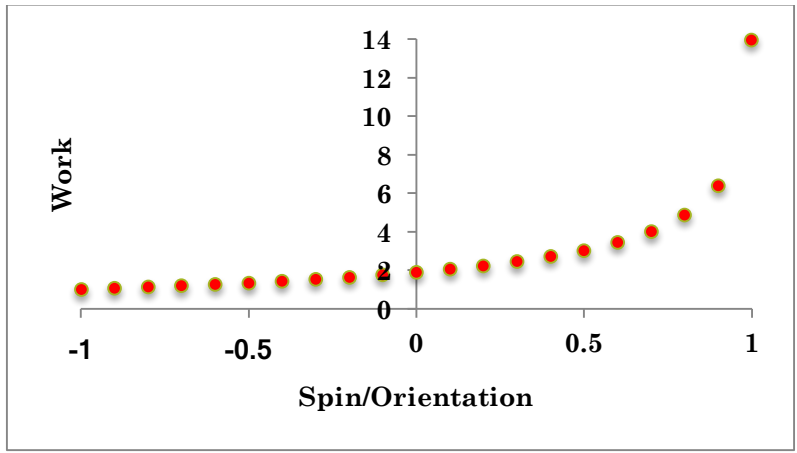

Figure 2: Work versus spin/orientation required to extract the angular momentum necessary to accrete material in a thin disk that extends to 30 gravitational radii normalized to the value at high retrograde spin.

3. Implications for numerical simulations

In the previous section we showed that the fraction of conserved energy in the electromagnetic field needed to produce a given accretion rate is always larger in the prograde regime compared to the retrograde case. Because prograde disks have a greater amount of total energy available than retrograde ones, the fact that a greater amount of energy needs to be placed in electromagnetic fields does not by itself imply a difference between retrograde and prograde disks. However, the fact that prograde accretion disks live or occupy a region of the gravitational potential that is different from the retrograde disks, produces an important difference, one that has not yet been identified. The metric term associated with proper distance in the radial direction of the equatorial plane of the disk in Boyer-Lindquist coordinates is

$$g_{rr} = r^2/(r^2 - 2Mr + a^2)$$

where $r$ is the radial distance, M is the black hole mass, and $a$ is the dimensionless spin. While such a metric term does not discriminate between prograde and retrograde disks, the fact that prograde inner disk edges for high black hole spin approach the radial distance of 1.23 $GM/c^2$ while high spin retrograde disks are located at inner edges approaching the radial value 9$GM/c^2$, implies that proper distances between magnetically connected regions required to extract angular momentum, increase for prograde disks. This constitutes a basic asymmetry between retrograde and prograde disks that forces energy to be distributed in disks differently in order to accomplish the task of accreting material. In other words, the mechanism for angular momentum extraction in accretion disks forces the prograde disk to place a relatively greater amount of their electromagnetic energy into radially connecting plasma in the disk plane to compensate for the fact that proper distances are greater. Therefore, the fact that a greater amount of energy is available for more prograde disks, does not seem to allow prograde disks to be more effective, or even as effective, in extracting their angular momentum as retrograde disks. The increase in the energy is precisely compensated by the fact that greater radial distances must be connected for the MRI to operate. In short, there are two aspects of disk orientation that matter. The first is that for fixed radial locations such as from 10$GM/c^2$ to any other location further out in the disk (i.e for regions of the disk that are part of accretion disks that are either prograde or retrograde), retrograde disks require less work to extract their angular momentum than prograde disks. If the inner and outer disk radii were independent of orientation, therefore, our analysis shows that retrograde disks are favored energetically in their ability to provide more energy in large scale electromagnetic fields than prograde disks. The second involves the failure to circumvent this result by appealing to an additional reservoir of energy for prograde disks. Additional energy is available to prograde disks because they live in regions of higher gravitational potential, with the hope being that such additional reservoir of energy could be used to increase the amount of energy allotted to the large scale electromagnetic field. However, our analysis suggests that the mechanism for angular momentum extraction (i.e. the MRI) requires that the additional energy go into electromagnetic fields in the disk plane. In fact, because connections that establish the MRI produce radially longer field connections for prograde disks living closer to high spinning black holes, magnetic reconnection phenomena will become increasingly more important for disks near black holes. This suggests that high spinning prograde disks will struggle more than our analysis suggests in converting energy into large scale electromagnetic fields.

Our analysis shows the effectiveness of angular momentum extraction and thus accretion in black hole systems to depend crucially on disk orientation in a basic way. There is, in other words, an orientation symmetry-breaking effect in black hole accretion. Retrograde disks are relatively better suited for extracting their angular momentum and this can also enhance magnetic fields near the event horizon (Mikhailov et al 2015).

Note that some GRMHD simulations do indeed produce lower accretion rates in prograde configurations (McKinney & Gammie 2004). Despite the fact that simulations seem to track the analytic constraints, they still produce greater black hole-threading magnetic fields in the prograde direction, thereby violating our energy conservation constraints, resulting in stronger jets. While we cannot converge on the exact reason from first principles, the likely explanation for this violation may concern the following approximations. First, as discussed above, the electromagnetic field is assumed in the initial configuration, and its strength is spin-independent. This is true whether the initial electromagnetic configuration involves closed loops within the disk proper as in the earlier simulations (e.g. Hirose et al 2004; McKinney & Gammie 2004) or loops that lead to the magnetically flooded black holes as in recent MADs (e.g. McKinney, Tchekhovskoy & Blandford 2012). As a result of this, all the work that the accretion disk would have done to produce that specific accretion rate, is assumed to occur without any of the conserved energy going into the production of magnetically connected, spatially separated, regions, in the disk. Therefore, all of that unconsumed energy can now be turned into a large-scale field by disk dynamo action. Second, the black hole spin dependence of the disk dynamo – the process that takes the disk field and turns it into large-scale field – is absent. In fact, in GRMHD there is no equation of the form

$$J^a = \sigma F^{ab} U_b \qquad (9)$$

with $J^a$ the current 4-vector, $\sigma$ the large but finite conductivity, $F^{ab}$ the (2,0) Faraday tensor, and $U_b$ the velocity one-form. Instead, $\sigma$ is assumed to be infinite-valued, and the finite current density results from the assumption of zero proper electric fields according to

$$F^{ab} U_b = 0, \qquad (10)$$

which, is referred to as the 'ideal Ohm's law'. Note that equation (9) actually constitutes a violation of special relativistic causality since it legislates an instantaneous current in reaction to the fields (Koide 2008). The exact impact of this remains poorly explored. In a sense, therefore, GRMHD does worse by implementing equation (10), which also, of course, implies in principle that magnetic reconnection is absent. However, numerical diffusion comes to the rescue in some respect by providing for dissipation of the electromagnetic field via turbulence, which is obviously a violation of (10). Does the numerical failure to uphold (10) imply strict compliance with (9)? That would surely be impossible. And, importantly, the black hole spin dependence in (9) is absent in GRMHD. The claim in GRMHD, however, is that the black hole spin dependence in (9) that is absent in GRMHD is a higher order effect and therefore negligible and that turbulence effectively subsumes the relevant physics without having to resort to (9) (see Lazarian et al 2015 for a review). But even in the context of MHD turbulence mediated by the MRI, the chaotic non-linear process cannot violate the underlying conservation principles. The importance of subgrid physics in the context of a non-ideal Ohm's law is emphasized in Bucciantini & Del Zanna (2013), where a more generalized general relativistic Ohm's law appears.

The possibility that energy is not properly treated or accounted for in GRMHD simulations is not new. Meier (2012 p. 700) documents the lack of energy conservation along a magnetic field line in the simulations of McKinney (2006), noting that the energy in matter and electromagnetic field does not remain constant. In the notation used by McKinney (2006), the total energy at infinity is split into the matter part (superscript MA) and the electromagnetic part (superscript EM)

$$\mu = \gamma_\infty = \gamma^{MA}{}_\infty + \gamma^{EM}{}_\infty. \tag{11}$$

While Meier's concern surrounds the physics of recollimation and the accounting of the energy required for that process, he emphasizes that not only does the total energy drop by two orders of magnitude within only 100 gravitational radii, but that the nature of this loss, i.e. where the energy goes, is unclear. But this, as Meier points out, is perhaps not surprising since the best simulations resolve regions that extend hundreds of meters for 10 solar mass black holes and much larger for millions to billion solar mass black holes. Energy redistribution on such scales cannot therefore be properly treated and a kind of averaging must be administered that allows the energy lost numerically to be reinserted in the simulation, reintroduced as heat despite the absence of the microphysics that physically accomplishes this. Because accurate Riemann solvers are computationally expensive, most simulations adopt more approximate methods that are also more diffusive (see White & Stone 2016 for a discussion of this).

The combination of the poor accounting of energy in GRMHD together with the constraints from Noether's theorem, suggest that the implementation of GRMHD simulations is responsible in some way for smuggling in or out, in a spin-dependent way, a source of energy that is not compatible with the physics of black hole accretion.

4. Conclusions

We have argued from the perspective of conservation laws in general relativity that electromagnetic fields threading black holes that are brought to the hole by an accretion disk are subject to constraints that emerge from the way that accretion operates, namely magnetically connected, spatially separated regions. As a result of this, we have argued that not only are accretion rates expected to be lower in the prograde regime (as is in fact seen in GRMHD), but that prograde disks likely accrete efficiently at the expense of their large-scale electromagnetic fields. In other words, energetic processes contingent on the strength of large-scale electromagnetic fields will tend to be weaker for prograde accreting black holes. Prograde accretion mediated by the MRI instability requires relatively more of the total available energy to go into magnetic energy in the accretion disk (into radial and azimuthal field components), leaving less for the generation of large-scale electromagnetic fields.

Despite not being able to pinpoint the exact breakdown implicit in the incompatibility between GRMHD and the results of section 2, we have explored two possibilities. The first is that the initial conditions adopted for the electromagnetic field in the disk may amount to a violation of the transport or creation of the field in-situ. And, the second, amounts to a claim that turbulent diffusion and resistivity generated numerically may be washing away important black hole spin dependent microphysics. While we have motivated our results from first principles, a thorough understanding of this dynamics requires detailed analysis of the complex astrophysical transfer of energy from particles to fields, including magnetic reconnection physics in non-ideal MHD or even beyond the MHD simplification, most of which is currently beyond the state-of-the art in numerical simulations of black hole accretion.

Acknowledgments

I thank Professor Yuri N. Gnedin and two anonymous referees for their contribution to the clarity of the paper and for a crucial insight into total electromagnetic energy considerations that had not been highlighted in my original submission.